\documentclass[a4paper, 12pt]{article}
\usepackage{amsmath}
\usepackage{amssymb}

\begin{document}

\title{A MHD invariant and the confinement regimes in Tokamak}
\author{F. Spineanu and M. Vlad}
\date{}
\maketitle

\begin{abstract}
Fundamental Lagrangian, frozen-in and topological invariants can be useful
to explain systematic connections between plasma parameters. At high plasma
temperature the dissipation is small and the robust invariances are
manifested. We invoke a frozen-in invariant which is an extension of the
Ertel's theorem and connects the vorticity of the large scale motions with
the profile of the safety factor and of particle density. Assuming
ergodicity of the small scale turbulence we consider the approximative
preservation of the invariant for changes of the vorticity in an annular
region of finite radial extension (i.e. poloidal rotation). We find that the
ionization-induced rotation triggered by a pellet requires a reversed-$q$
profile in an off-axis region of the core. In the $H$-mode, the invariance requires the accumulation of the
current density in the rotation layer at the edge. Then this becomes a vorticity-current
sheet which may explain experimental observations related to the penetration
of the Resonant Magnetic Perturbation and the filamentation during the Edge
Localized Modes.
\end{abstract}

\section{Introduction}

To understand and predict properties of the tokamak plasma it is necessary
to investigate turbulent-induced transport processes and, of equal
importance, coherent structures (like zonal flows, H-mode rotation layers
and convective cells). The analytical and numerical approaches and the
methods are however largely different, from statistical theory to exact
integrability. It is then helpful to remember that both physical aspects,
even if different in their manifestation, must obey a basic set of
constraints which are due to the existence of Lagrangian invariants and of
frozen-in invariants. The invariants are constraints that must be verified
by any dynamical evolution of the system. In general for real plasma (\emph{%
i.e.} with finite resistivity and viscosity) the invariance property can
only be approximative, but the high temperature (typical for fusion reactor)
reduces the collisional effects even in regions close to the plasma edge.
Then the persistence of some invariants can still be used as a benchmark for
our analytical models and for numerical simulations, similar to the
conservation of the energy or the angular momentum.

We will be interested in the invariants that include in their expression the
plasma vorticity $\mathbf{\omega }=\mathbf{\nabla \times v}$, \emph{i.e.}
the sheared plasma rotation, which has constantly been proven an important
factor in the confinement. In the specific case of the tokamak, variation of
the parameters composing the invariant appear as small perturbations around
the high cyclotron frequency $\mathbf{\Omega }_{c}$, which is a kind of
\textquotedblleft condensate of vorticity\textquotedblright . Although it is
a passive component of the \textquotedblleft absolute
vorticity\textquotedblright\ $\mathbf{\omega +\Omega }_{c}=\mathbf{\nabla
\times }\left( \mathbf{v+}\frac{1}{2}\mathbf{\Omega }_{c}\times \mathbf{r}%
\right) $ the high magnitude of $\Omega _{c}$ reduces the practical
relevance of the invariant to those situations where the physical vorticity
is sufficiently high. We expect that the $H$-mode, the zonal flows and the
internal transport barrier belong to this class.

If the Lagrangian and the frozen-in invariants are difficult to be tested
directly, they are however responsible for another class of invariants, of
topological nature.

\bigskip

\section{Geometry versus plasma dynamics, reflected in types of invariants}

\subsection{Topological invariants}

Below we underline the difference between the topological invariants and
those that involve fluid advection.

There is a hierarchy of topological invariants \cite{arnold1998} \cite%
{akhmetievruzmaikhin}, the lowest and most familiar being the Gauss linking
of two lines in space, a discrete number showing how many times one line
turns around the other. If the Kolmogorov length is a local measure of the
chaotic state (\emph{i.e. }the presence of the stochastic instability), the
Gauss linking number is a non-local variable which can also be stochastic 
\cite{tanaka1}, \cite{tanaka2}, \cite{flmadi17}. It is a quantitative
measure that goes beyond the Chirikov criterion of overlapping of chains of
magnetic islands developing at nearby rational surfaces, a criterion usually
invoked for the onset of magnetic stochasticity \cite{flmadi18}, \cite%
{brereton1}. More generally, the lines in space can be magnetic field lines
or streamlines of the plasma flow. The fluid quantities that are associated
are the magnetic helicity whose volume density is $\mathbf{A\cdot B}$,
kinetic helicity $\mathbf{v\cdot \omega }$ and cross-helicity $\mathbf{%
v\cdot B}$ \cite{moffatt1978magnetic}. The connection between the helicity
that is defined by the density of a continuous field in the fluid volume and
the discrete Gaussian link numbers \cite{khesinwendt} can be seen in the
following way. Consider a set of lines $\gamma _{k}$ that are closed, like
in tokamak, or possibly after extending to infinity. We take as basic fields
defined in the volume of the fluid/plasma, the magnetic potential $A_{\mu
}=\left( A_{k},A_{0}\right) $ and the velocity $v_{\mu }=\left(
v_{k},c_{0}\right) $ (where $c_{0}$ is a constant) and define the
circulation as the integrals of the potential or velocity along these lines 
\begin{equation*}
\oint_{\gamma _{k}}\mathbf{A\cdot dl}\ \ \text{and}\ \ \oint_{\gamma _{k}}%
\mathbf{v\cdot dl}
\end{equation*}%
As results from the Stokes theorem, these are the fluxes of the magnetic
field $\mathbf{B=\nabla \times A}$ respectively of the vorticity $\mathbf{%
\omega =\nabla \times v}$ through the surface bounded by the curve $\gamma
_{k}$. The freezing-in of $\mathbf{B}$ and respectively the Kelvin theorem
ensure that, when the line $\gamma _{k}$ is carried by the flow (\emph{i.e.}
it is a \emph{material} curve) the fluxes remain constant. Consider the
variable 
\begin{equation*}
\exp \left( \sum\limits_{k}\oint_{\gamma _{k}}A_{\mu }dx^{\mu }\right)
\end{equation*}%
(at fixed time $x^{0}$ the integrand is identical with that of the previous
expression; all calculations can be done for $v_{\mu }$ as well) for a
particular configuration of the field $A_{\mu }$ and calculate the average
of this quantity over an ensemble of realizations of the field $A_{\mu }$
defined in space $\mathbf{R}^{3}$. The statistical accessibility of a state
with flux $\oint_{\gamma _{k}}A_{\mu }dx^{\mu }$ is given by the
Boltzmann-like weight factor, defined as the exponential of the total
helicity in the volume%
\begin{equation*}
\exp \left( -\int d^{3}x\varepsilon ^{\mu \nu \rho }A_{\mu }\partial _{\nu
}A_{\rho }\right)
\end{equation*}%
The integrand is the general form of the usual helicity $\mathbf{A\cdot B}=%
\mathbf{A\cdot }\left( \mathbf{\nabla \times A}\right) $. This Boltzmann
factor "penalizes" states with high total amount of helicity in the volume
by making them less accessible. It can be shown that the functional integral 
\begin{equation*}
\int \mathcal{D}\left[ A_{\mu }\right] \exp \left(
\sum\limits_{k}\oint_{\gamma _{k}}A_{\mu }dx^{\mu }\right) \exp \left( -\int
d^{3}x\varepsilon ^{\mu \nu \rho }A_{\mu }\partial _{\nu }A_{\rho }\right)
\end{equation*}%
which is the statistical average of the exponential of the field fluxes
through surfaces bounded by the fixed set of closed spatial curves $\gamma
_{k}\ \left( k=1,2,...\right) $ , over fields of all possible helicities, is
the exponential of the sum of Gaussian linking number of pairs of curves $%
\gamma _{k}$. The \emph{entanglement} of $\gamma _{k}$'s is a purely
geometric characteristic and for ideal fluids is conserved by the motion.
Higher order topological invariants can be found, with more sofisticated
definitions \cite{akhmetievruzmaikhin}. The nonconservation of the
topological content of the magnetic field configuration is determined by
reconnection. In the tokamak conditions, only the presence of dissipation
allows breaking-up and reconnections of magnetic field lines, thus making
possible transitions between classes of states with distinct topological
content. When the fluid is slightly non-ideal, low-order topological
invariants like the \emph{entanglement} are robust and they still remain
relevant, while higher order topological invariants are gradually lost \cite%
{isichenko1}, \cite{KraichnanMontgomery}. This is better illustrated by the
two-dimensional topology of streamlines of a quasi-ideal fluid. The inverse
cascade in $2D$ leads asymptotically to a highly coherent state of the flow,
a simple dipole in the double periodic plane domain, after all positive
respectively negative vorticity elements have been separated and
concentrated \cite{Montgomery1992}. This evolution essentially traverses a
large set of distinct topological configuration and this is possible only
through break-up and reconnection of the streamlines (\emph{i.e.} violation
of the Kelvin invariance), made possible by the finite dissipation. The
dissipation is located in very small volumes where the gradients of the
magnetic field (respectively vorticity for fluids) are very large and the
ensemble of these points is fractal in the total volume \cite{Levich1987}.
The amount of energy lost at every reconnection is small and for this reason
the total energy is still a good invariant while the topology undergoes
radical changes \cite{flmadichaos}, \cite{FlorinMadiXXX2013}.

We note from the example given above that a topological invariant does not
involve directly the dynamics of fluid motion. The topological properties
mentioned above are commonly associated to sets of lines with clear physical
meaning (magnetic lines and lines of fluid flow) whose stochastic
entanglement is the cause of high rate particle and heat transport. Other
sets of lines must be considered, which can be defined according to the same
method as for the magnetic field. The Clebsch representation of the magnetic
potential $\mathbf{A=\nabla }\theta +\alpha \mathbf{\nabla }\beta $ shows
that the lines of the magnetic field $\mathbf{B}=\mathbf{\nabla }\alpha
\times \mathbf{\nabla }\beta $ are defined by the intersection of two
surfaces $\alpha =$const, $\beta =$const. More generally, the two functions $%
\alpha $ and $\beta $ are Lagrangian invariants that define the frozen-in
invariant $\mathbf{B/}\rho $. A comprehensive explanation is presented by
Tur and Yanovskii \cite{turyanovsky1993}. The purpose of the above
discussion is to underline the distinction between the topology-preserving
constraints and the fluid invariants, the latter being our subject.

\subsection{Invariants of the plasma dynamics}

The Lagrangian invariants are quantities $I$ that remain constant along the
lines of flow%
\begin{equation}
\frac{\partial I}{\partial t}+\left( \mathbf{v\cdot \nabla }\right) I=0
\label{eq1}
\end{equation}%
An elementary example is the density $\rho $. We can see the flow as a
collection of streamlines each carying as unique label the initial point $%
\mathbf{x}_{0}=\mathbf{x}\left( t=0\right) $. This characteristic remains
the same for all moments of displacement of a particle\ along the streamline
and so $\mathbf{x}_{0}$ is a Lagrangian invariant. The integrals $\mathbf{J}$
defined by the equation%
\begin{equation}
\frac{\partial \mathbf{J}}{\partial t}+\left( \mathbf{v\cdot \nabla }\right) 
\mathbf{J=}\left( \mathbf{J\cdot \nabla }\right) \mathbf{v}  \label{eq2}
\end{equation}%
and are vector fields that are advected by the flow as frozen-in. For
example, the magnetic field $\mathbf{B}$ is frozen into the plasma.
Frozen-in quantities can be obtained from two Lagrangian invariants $I_{1}$
and $I_{2}$, as%
\begin{equation}
\mathbf{J}=\frac{1}{\rho }\mathbf{\nabla }I_{1}\times \mathbf{\nabla }I_{2}
\label{eq3}
\end{equation}%
which is suggested by the fact that $I_{k}\left( x,y,z,t\right) =$ const
along the streamlines defines a surface and the gradient of the scalar
invariant is proportional with the vector of the element of area 
\begin{equation}
\mathbf{\nabla }I_{k}\sim \rho d\mathbf{S}  \label{eq4}
\end{equation}%
Then $\mathbf{\nabla }I_{1}\times \mathbf{\nabla }I_{2}$ is the line of
intersection of the two surfaces $I_{1,2}=\ $const. More generally,
multiplying this expression with another Lagrangian invariant, 
\begin{equation}
J_{\alpha }=\frac{1}{\rho }\varepsilon _{\alpha \beta \gamma }\varepsilon
_{ijk}I_{i}\frac{\partial I_{j}}{\partial x^{\beta }}\frac{\partial I_{k}}{%
\partial x^{\gamma }}  \label{eq5}
\end{equation}%
one obtains a new class of frozen-in invariants. We note each Lagrangian
invariant is a scalar $I_{k}$ while $\rho \mathbf{J}$ is a current in the $%
\left( 3+1\right) $ - dimensional geometry. For the two-dimensional ideal
fluid the ratio 
\begin{equation}
\mathbf{J}^{\omega }=\frac{\mathbf{\omega }}{\rho }  \label{eq9}
\end{equation}%
is a frozen-in invariant. Since $\mathbf{\omega }=\mathbf{\nabla \times v}$
we have $\mathbf{\nabla \cdot }\left( \mathbf{J}^{\omega }\rho \right) =0$.
For incompressible fluid $\mathbf{\nabla \cdot v}=0$, $\rho =$ const and the
equation 
\begin{equation}
\frac{d}{dt}\left( \rho \mathbf{J}^{\omega }\right) =0  \label{eq10}
\end{equation}%
becomes the Euler equation. For the plasma immersed in strong magnetic field
(similarly for planetary atmosphere, both being well approximated by a
two-dimensional geometry, see \cite{Pedlosky}) the invariant (Ertel) is 
\begin{equation}
\mathbf{J}^{E}=\frac{\mathbf{\omega }+\mathbf{\Omega }_{c}}{n}  \label{eq11}
\end{equation}%
where $\mathbf{\Omega }_{c}=\Omega _{c}\widehat{\mathbf{e}}_{z}$ is the
cyclotron frequency (in magnetized plasma) or the Coriolis frequency (in
planetary atmosphere); $\widehat{\mathbf{e}}_{z}$ is the unit vector
perpendicular on the plane. From this invariant it is derived the
Charney-Hasegawa-Mima (quasi-three dimensional) equation, governing the
turbulent fluctuations and the cuasi-coherent vortical structures at the
level of the ion Larmor radius \cite{hasegawamima}. If the gyration
frequency is far from the frequency scale associated to the plasma motions,
a natural extension of this invariant exists for charged particles
(separately for electrons and for ions), which includes the absolute
vorticity and the magnetic field \cite{nonlinearmir}. It is a frozen-in
invariant of the same structure as Eq.(\ref{eq2})%
\begin{equation}
\frac{d}{dt}\frac{\mathbf{\omega +}\eta \mathbf{B+\Omega }_{c}}{\rho }%
=\left( \frac{\mathbf{\omega +}\eta \mathbf{B+\Omega }_{c}}{\rho }\cdot 
\mathbf{\nabla }\right) \mathbf{v}  \label{eq13}
\end{equation}%
where $\eta $ is a dimensional constant, $\eta =\left\vert e\right\vert /m$.
Since we are interested in plasma motion we will use this invariant with
reference to ions. Eq.(\ref{eq13}) is the extended form of the Ertel's
theorem which includes the magnetic field. The lines of force of the field%
\begin{equation}
\mathbf{J=}\frac{\mathbf{\nabla \times v+}\eta \mathbf{B\mathbf{+\Omega }_{c}%
}}{n}  \label{eq15}
\end{equation}%
are strongly influenced by the turbulence. We can choose an arbitrary point $%
\mathbf{x}_{1}$ situated on a line of force of $\mathbf{J}$ and the value $%
\mathbf{J}\left( \mathbf{x}_{1}\right) $ taken in that point. Assuming that
the turbulence is ergodic another line of force, originating from a
different initial position $\mathbf{x}_{2}$ will come close to the chosen
point. Since the two lines of force are carrying different values of $%
\mathbf{J}$ (as determined by their initial conditions) the real value that
will be "measured" is an average taken over a small spatial patch around $%
\mathbf{x}_{1}$, since the viscosity will smooth out the strong differences.
The space-time evolution of $\mathbf{J}$ initialized at $\mathbf{x}_{1}$,
requires the fields $\left( \mathbf{\omega ,B,}n\right) $ that enter its
expression to adjust themselves such as to ensure invariance. If in a region
of plasma the vorticity is higher than that of $\mathbf{x}_{1}$ then the
magnetic field $\mathbf{B}$ and/or the density $n$ must compensate for this
change.

We note finally that in the invariant $\mathbf{J}$ are present the
\textquotedblleft relative\textquotedblright\ $\mathbf{\omega }$ and the
\textquotedblleft absolute\textquotedblright\ $\mathbf{\omega +\Omega }_{c}$
vorticities (in the language of the fluid physics).

\subsection{Brief comparison with the TEP invariants}

Invariants that explicitely constrain the motions in plasma have been
derived in three basic approaches: relabeling invariance within the
Hamiltonian formulation for fluids; analytical derivation of frozen-in and
Lagrangian invariants as properties of the basic equations; and Turbulent
Equipartition (TEP).

Our focus will be on the MHD invariant found by Sagdeev, Moseev, Tur and
Yanovskii (SMTY), Eq.(\ref{eq15}), which can be derived from the basic
two-fluid plasma equations. In increasing complexity, the SMTY invariant can
be seen as an extended form of well known frozen-in invariant $\mathbf{B}%
/\rho $. Or, there is a particularity of the latter invariant, in that it
can also be derived, in a similar form, within the concept of turbulent
equipartition (TEP). Then one can be tempted to suppose that the SMTY
invariant, or other frozen-in invariants, can also be connected somehow with
the TEP. We find useful to discuss this aspect since the two possible views
on the SMTY invariant would be substantially different and would have impact
on the applications. From the ideal MHD model%
\begin{eqnarray}
\frac{\partial \mathbf{v}}{\partial t}+\left( \mathbf{v\cdot \nabla }\right) 
\mathbf{v} &=&-\frac{1}{\rho }\mathbf{\nabla }p+\frac{1}{\rho }\frac{\mathbf{%
\nabla \times B}}{\mu _{0}}\times \mathbf{B}  \label{eq1178} \\
\frac{\partial \mathbf{B}}{\partial t}+\left( \mathbf{v\cdot \nabla }\right) 
\mathbf{B} &=&\left( \mathbf{B\cdot \nabla }\right) \mathbf{v}  \notag \\
\frac{\partial \rho }{\partial t}+\left( \mathbf{v\cdot \nabla }\right) \rho
&=&0  \notag
\end{eqnarray}%
one obtains%
\begin{equation}
\frac{\partial }{\partial t}\left( \frac{\mathbf{B}}{\rho }\right) +\left( 
\mathbf{v\cdot \nabla }\right) \left( \frac{\mathbf{B}}{\rho }\right)
=\left( \frac{\mathbf{B}}{\rho }\cdot \mathbf{\nabla }\right) \mathbf{v}
\label{eq1179}
\end{equation}%
and this shows that $\mathbf{B}/\rho $ is a frozen-in \emph{fluid} invariant.

Unlike the fluid MHD the TEP essentially relies on the particle dynamics.
One starts from the invariants of the equations of motion of a single
charged particle: $\left( \epsilon ,\mu ,J\right) $, which are derived in
the particular geometry of the toroidally magnetic confinement. This is much
less general than the MHD equations above. Since the free motion of the
particle is periodic (gyration, bounce on banana orbit, toroidal precession)
one can introduce action-angle variables \cite{tepisigrdiaprl}. The
interaction with the turbulence leads to the loss of some of the invariants,
if the frequency of periodic motion is less than the typical frequency of
the electrostatic turbulence \cite{tepisigrdiayush}. Due to the high
gyration frequency the magnetic momentum $\mu $ remains invariant $\mu
=v_{\perp }^{2}/\left( 2B\right) $. \ The phase space can be reduced to $2D$ 
\emph{i.e.} $\left( x,y,v_{x},v_{y}\right) $ where the invariant $\mu =\mu
_{0}$ defines a hypersurface. The velocity space in the laboratory frame $%
\left( v_{x},v_{y}\right) $ is mapped onto $\left( \mu ,\zeta \right) $
where $\zeta $ is the gyrophase. The turbulence produces uniform
distribution on the hypersurface $\mu =$ const of the probability to find
the state of the particle in some infinitesimal $2D$ volume. We need to
express the result in the real-space coordinates which means to return from $%
\left( \mu ,\zeta \right) $ to $\left( v_{x},v_{y}\right) $. The density
which is uniform on $\mu =$ const is multiplied by the Jacobian of the
transformation and this introduces the dependence of the parameters \cite%
{tepnycyankov}%
\begin{equation}
n\left( x,y\right) \sim \frac{\partial \left( v_{x},v_{y}\right) }{\partial
\left( \mu ,\zeta \right) }=B  \label{eq1180}
\end{equation}%
or $n/B=$const. This result is identical to the MHD result $B/\rho =$const
(after reduction to $2D$ there is no need to exhibit the vectorial nature of 
$B$). We note that in this TEP derivation the density $n$ occurs as the
distribution function integrated over the velocity space. This is because we
know the density on the sub-manifold determined in general by constants of
motion of the particle, like $\mu =$ const., more precisely we know that
this is constant. And we want to carry back in the laboratory frame this
result. The magnetic field $B$ appears from the Jacobian (which is the ratio
between two volume elements) of the transformation between the two systems
of coordinates: canonical and laboratory. The physical meaning of TEP
restricts the effect to trapped electrons \cite{tepnycyankov} and expects to
extend to circlating electrons and to ions through collisions.

The fact that the frozen-in invariant is also derived in TEP suggests to
explore similarities for $\mathbf{\omega }/\rho $ (for fluids) and $\left( 
\mathbf{\omega +\Omega }_{ci}\right) /\rho $ (for plasma). The latter
invariant is derived for tokamak plasma starting from basic equations \cite%
{hasegawamima} and in the general case from relabeling invariance \cite%
{salmonhamilt}.

\section{The large scale vorticity in tokamak}

Usually one encounters the vorticity as an essential field in the dynamics
of the drift wave instabilities and turbulence, \emph{i.e.} at spatial
scales of the order of $\rho _{s}$ (millimeter). In these cases one has $%
\omega =\Delta \phi $ where $\phi $ is the fluctuating electric potential of
a drift-type instability. The vorticity that is involved in the invariant
Eq.(\ref{eq11}) is a large scale variable, resulting from the sheared
rotation velocity, either poloidal or toroidal. It refers to flows on
spatial scales of the order of the minor radius. The vorticity $\omega $,
which is the radial derivative of the sheared velocity profile, can be
localized if there are fast variation of $v\left( r\right) $ on the minor
radius, as is the case for the rotation layer in the H-mode or for flows
associated to narrow Internal Transport Barriers.

The importance of the Ertel's theorem asserting the invariance in Eq.(\ref%
{eq11}) is well known. The $2D$ compressibility of the ion polarization
drift and the assumed Boltzmann distribution of density perturbation along
the magnetic field lines lead to a single equation for the electric
potential in the two-dimensional (poloidal plane) approximation of the drift
waves, \emph{i.e.} the Charney-Hasegawa-Mima equation. The invariant Eq.(\ref%
{eq15}) , which can be seen as an extension of the Ertel's theorem, is
particularly important in a plasma where the zonal flows, the H-mode layer
of plasma rotation and the Internal Transport Barriers imply spatial
variation of the magnitude of plasma velocity, \emph{i.e.} vorticity. When
an element of plasma that initially has a \ set of values $\left( v_{\theta
},B_{\theta },n\right) ^{\left( 0\right) }$ is driven by the irregular
motions of drift turbulence in a different region, where a different
magnitude $v_{\theta }^{\prime }$ is imposed by the plasma rotation, the
magnetic field (or, profile of current density) and/or density must change,
to keep $J_{z}$ invariant.

Physical sources of generation of poloidal and toroidal rotation are
Reynolds stress (sustained by turbulence), Stringer mechanism, loss of ions
to the limiter, neoclassical intrinsic ambipolarity and, in addition,
injection of torque from external sources (NBI, ICRH) or from creation of
alpha particles by fusion reactions. Since all have rates with spatial
variation, which is reflected in spatial variation of the velocity, the
presence of vorticity is ubiquituous. Therefore the invariance Eq.(\ref{eq15}%
) should be examined in the context of the mentioned sources of vorticity.

In the present work we include another source of torque (and rotation),
arising from an equally ubiquituous phenomenon in the tokamak plasma, the
ionization of neutral atoms \cite{florinmadidensity}.

Every event of ionization of a neutral particle in tokamak plasma it
followed by the displacement of the newly born charges (electron and ion)
towards the equilibrium neoclassical orbits. The motion of the "new" ion on
either circulating and trapped (banana) orbit is asymptotically periodic.
However there is the first interval, just after the ionization, when the ion
evolves to take the periodic trajectory, which is an intrinsically
non-periodic transient. This displacement is an effective radial current. At
the end of this finite, transitory part, the motion becomes periodic and
there is no radial current. The short and unique transient radial current,
multiplied by the rate of ionization, can produce a significat torque and
plasma rotation, thus influencing the confinement. We introduce the rate of
ionization $\overset{\cdot }{n}^{\left( 0\right) }S\left( x,t\right) $ where
the first factor is an average magnitude ($ions/m^{3}/s$) and $S\left(
x,t\right) $ is a function of order unity that represents the space
inhomogeneity and time non-uniformity of the rate of ionization. The total
current traversing the point $x$ at time $t$ is obtained by summing over all
the individually short currents%
\begin{eqnarray}
j\left( x,t\right) &=&\overset{\cdot }{n}^{\left( 0\right)
}\int_{0}^{t}dt_{0}\int_{0}^{x}dx_{0}\ S\left( x_{0},t_{0}\right)
\label{eq16} \\
&&\times \Theta \left( t_{\Delta }-t\right) \Theta \left( t-t_{0}\right)
\times \left\vert e\right\vert v_{D}\delta \left( x-v_{D}t\right) \times
\Theta \left( x-x_{0}\right)  \notag
\end{eqnarray}%
Here $\left( x_{0},t_{0}\right) $ are the place and time of creation of a
new ion, $v_{D}$ is the neoclassical drift velocity and $t_{\Delta }$ is the
time of transit of the ion from its creation to the "center" of the periodic
orbit, \emph{i.e.} the time that there is an electric current. The distance
on which the elementary contribution is non-zero is $\Delta =v_{D}t_{\Delta
} $. The Heaviside functions $\Theta $ underline the limited time of
existence of a single contribution to the current.

The interval of existence of the current of a single ion is temporarly short
and is spatially small. Then we can expand $S\left( x_{0},t_{0}\right) $ and
after integration over the nearby contributions we obtain the expression of
the radial current%
\begin{equation}
j\left( x,t\right) \approx -\frac{1}{2}\left\vert e\right\vert \overset{%
\cdot }{n}_{0}^{ioniz}\left( \frac{\partial S}{\partial x}\right) \rho
_{i}^{2}q^{2}\varepsilon ^{-1/2}  \label{jtapp}
\end{equation}%
The details of this calculation are presented elsewhere (\cite%
{florinmadidensity}). This radial current interacts with the magnetic field
producing a torque that acts on the new ions. Since rough estimations show
that, in some particular situations, the magnitude of the torque is
comparable or higher with the Transit Time Magnetic Damping we will focus on
the poloidal rotation, which has a strong effect on the turbulence.

\section{A contribution to the reversed $q$ profile}

The pellet enhanced performance (PEP) experiments \cite{pepjetreview}, \cite%
{Hugon} have shown the formation of strong pressure gradients that drive a
bootstrap current peaked off-axis. The bootstrap current depends essentially
on collisions but within the known limits: electron-electron collisions are
necessary for the current of trapped electrons to be transferred to
circulating electrons, and electron-ion collisions saturate the bootstrap
current by momentum transfer to the ions \cite{bootstrapcordey}. At the high
temperature ($\sim 5\ KeV$) collisions are not frequent, $\nu _{ee}\sim
n/T_{e}^{3/2}$. It also depends on the presence of trapped particles \cite%
{bootstrapbickerton} which, close to the center, are less numerous $\sim 
\sqrt{r/R}$. We note in the following the the SMTY invariant suggests the
existence of an additional contribution that can enhance $j\left( r,t\right) 
$ beyond the bootstrap current. It comes from any local creation of plasma
sheared rotation, \emph{i.e.} from the creation of a local vorticity, $%
\omega =\partial v_{\theta }/\partial r$. Several sources are possible, with
different degrees of effectiveness. The ICRH modifies the width of the
trapped ion banana orbits and the implicit radial excursion of the virtual
center of the bouncing ion is a radial current. This produces a sheared
rotation \cite{icrhwhite}. In \cite{florinmadidensity} we draw attention to
an unexplored mechanism of generating rotation, through the radial current
resulting from transitory events when a particle changes from trapped to
circulating or inversely. Changing between two periodic motions, the virtual
center of the averaged position in one state is displaced to correspond to
the other state, which means a transitory radial displacement of a charge,
therefore a radial current. ICRH and also ECRH will show this effect and
must be considered sources of $\omega $. Finally, as mentioned before, the
ionization of a pellet generates a radial current that can be substantial.
None of these sources of $\omega $ should be ignored but in the following we
refer in principal to the ionization-induced rotation.

Returning to Eq.(\ref{eq15}) we will assume that the two-dimensional
reduction of $\mathbf{J}$, \emph{i.e.} transversal on the poloidal plane,
maintains the invariance. In this case, the invariance of $J_{z}$ appears as
an extension of the Ertel's theorem which yields the Hasegawa Mima equation 
\cite{hasegawamima}.

The magnetic field is%
\begin{equation}
B\approx \frac{B_{0}}{R}+\frac{1}{2}\frac{B_{0}}{R}\frac{\varepsilon ^{2}}{%
q^{2}}+...  \label{eq20}
\end{equation}%
Then the quantity normalized to $\Omega _{c}$ 
\begin{equation}
J_{z}=\frac{\frac{\omega }{\Omega _{ci}}+\frac{\varepsilon ^{2}}{2q^{2}}+2}{n%
}  \label{eq21}
\end{equation}%
is maintained by turbulence as an approximative constant over the area where
one can assume ergodicity.

We consider a case where a pellet reaches a deep region in the plasma and
the ionizations produce a significant radial current. Then the plasma
response, consisting of the return current generates a poloidal rotation
with radial variation, which means with vorticity $\omega =\partial
v_{\theta }/\partial r\sim \partial ^{2}S/\partial r^{2}$. The density has a
slower response and we neglect its radial nonuniformity. It results%
\begin{equation}
\frac{-\left\vert \omega \right\vert }{\Omega _{ci}}+\frac{\varepsilon ^{2}}{%
2q^{2}}\sim \text{const}\times n-2  \label{eq22}
\end{equation}

When $\left\vert \omega \right\vert $ increases on a zone which is off-axis $%
q$ must correspondingly be reduced in order the two terms (of opposite sign)
to compensate their change and the sum to remain approximately constant.
This means that $q$ will aquire a minimum in that region. For a pellet, the
radial variation of the ionization $S\left( r,t\right) $ profile can produce 
$\omega \sim 10^{6}\ \left( s^{-1}\right) $ \cite{florinmadidensity}. It
suggests that a sudden onset of sheared rotation on a radial interval will
produce a contribution that can lead to a reversed-$q$ profile. The
condition is that $\omega $ is opposite to the magnetic field. The neglect
of the changes of the density can only be justified by the fast processes
related to ionization, the slow changes of $n\left( r\right) $ having been
noticed in experiments \cite{weisen1}.

We note again that only high variations of $\left\vert \omega \right\vert $
can have a visible effect, due to the high \textquotedblleft vorticity
background\textquotedblright\ $\Omega _{ci}$.

\section{The density of the current in the edge H-mode layer}

In many cases the $L$ to $H$-mode transition is characterized by the fast
formation of a narrow layer of strongly sheared poloidal plasma rotation 
\cite{burrellfaster}. It has also been found that the density that invades
the border through the X point is ionized and partially trapped producing a
strong torque \cite{changxtransport}. The torque is enhanced by the loss of
some of the trapped ions. The fast generation of high $\omega $ suggests to
study the possible consequences imposed by the SMTY invariant%
\begin{equation}
\frac{\pm \left\vert \omega \right\vert }{\Omega _{ci}}+\frac{\varepsilon
^{2}}{q^{2}}\sim \text{const}\times n-2  \label{eq23}
\end{equation}%
We have a large increase in \emph{magnitude} of the vorticity $\left\vert
\omega \right\vert $ and in consequence $q$ must decrease. In general, at a
strong modification of $\omega $ there follows a strong modification of $q$. 
\begin{equation}
\frac{1}{q\left( r\right) }=\frac{R}{rB_{T}}B_{\theta }=\frac{R}{rB_{T}}%
\frac{\mu _{0}}{2\pi r}\int_{0}^{r}j\left( r\right) 2\pi rdr  \label{eq24}
\end{equation}%
We divide the radial interval $\left[ a-\delta ,a\right] $ at the edge, 
\emph{i.e.} a poloidal shell, and write%
\begin{eqnarray}
\int_{0}^{a}j\left( r\right) 2\pi rdr &\approx &\int_{0}^{a-\delta }j\left(
r\right) 2\pi rdr+\int_{a-\delta }^{a}j\left( r\approx a\right) 2\pi rdr
\label{eq25} \\
&\approx &I_{1}+j\left( r\approx a\right) 2\pi a\delta  \notag
\end{eqnarray}%
Here we have introduced the notation%
\begin{equation}
I_{1}=\int_{0}^{a-\delta }j\left( r\right) 2\pi rdr  \label{eq26}
\end{equation}%
and we note that, even if the current in the poloidal shell $\left[ a-\delta
,a\right] $ is significant, the current on the surface $\left[ 0,a-\delta %
\right] $ is the main part of the current and actually $I_{1}$ is not too
much different than the total current $I$. It is, of course, smaller, $%
I_{1}\lesssim I$. With this notation we have%
\begin{equation}
\frac{1}{q}=\frac{1}{\overline{q}\left( a\right) }+\frac{1}{\overline{q}%
\left( a\right) }\frac{2\pi a\delta }{I_{1}}j\left( r\approx a\right)
\label{eq27}
\end{equation}%
where%
\begin{equation}
\overline{q}\left( a\right) \equiv \left( \frac{R}{aB_{T}}\frac{\mu _{0}}{%
2\pi a}I_{1}\right) ^{-1}  \label{eq28}
\end{equation}%
is approximately the safety factor at the edge. It results 
\begin{equation*}
\frac{1}{q^{2}}=\frac{1}{\left[ \overline{q}\left( a\right) \right] ^{2}}%
\left[ 1+\frac{2\pi a\delta }{I_{1}}j_{a}\right] ^{2}
\end{equation*}%
We introduce a notation $j_{I}$ which is the density in the narrow shell $%
\left[ a-\delta ,a\right] $ in poloidal plane%
\begin{equation}
j_{I}\equiv \frac{I_{1}}{2\pi a\delta }  \label{eq29}
\end{equation}%
This is very large since the term looks like, formally, the full current $I$
would be concentrated in the layer $2\pi a\delta $. Then we can expand 
\begin{equation}
\Delta \left( \frac{1}{q^{2}}\right) \equiv \frac{1}{q^{2}}-\frac{1}{\left[ 
\overline{q}\left( a\right) \right] ^{2}}=2\frac{1}{\left[ \overline{q}%
\left( a\right) \right] ^{2}}\frac{j_{a}}{j_{I}}  \label{eq30}
\end{equation}%
or%
\begin{equation}
j_{a}\approx j_{I}\left[ \overline{q}\left( a\right) \right] ^{2}\frac{1}{2}%
\Delta \left( \frac{1}{q^{2}}\right)  \label{eq31}
\end{equation}%
On the other hand we have the equation derived from the invariant%
\begin{equation}
\Delta \left( \frac{1}{q^{2}}\right) \approx \frac{\Delta \omega }{\Omega
_{ci}\varepsilon ^{2}}  \label{eq32}
\end{equation}%
assuming that the density has a slower reaction to the rotation.%
\begin{equation}
j_{a}\approx j_{I}\frac{\frac{1}{2}\Delta \left( \frac{1}{q^{2}}\right) }{%
\left( \frac{1}{\left[ \overline{q}\left( a\right) \right] ^{2}}\right) }%
=j_{I}\frac{1}{2}\frac{\Delta \omega }{\Omega _{ci}\varepsilon ^{2}}\left[ 
\overline{q}\left( a\right) \right] ^{2}  \label{eq33}
\end{equation}

For an order of magnitude-estimate we take a variation of the vorticity $%
\Delta \omega \sim \delta v_{\theta }/\delta r\sim 10\times 10^{3}\ /0.01\
\left( V/Tm^{2}\right) =10^{6}\ \left( s^{-1}\right) $ and $\Omega
_{ci}\varepsilon ^{2}\sim 10^{8}\times 0.1=10^{7}$. Then, for $\overline{q}%
\left( a\right) \approx q\left( r\sim a\right) $ such that $\left[ \overline{%
q}\left( a\right) \right] ^{2}\approx 20$, we have 
\begin{equation}
\frac{1}{2}\frac{\Delta \omega }{\Omega _{ci}\varepsilon ^{2}}\left[ 
\overline{q}\left( a\right) \right] ^{2}\sim 1  \label{eq34}
\end{equation}%
and from Eq.(\ref{eq33}) 
\begin{equation}
j_{a}\sim j_{I}=\frac{I_{1}}{2\pi a\delta }  \label{eq35}
\end{equation}%
which is very large, a factor of $a/\delta \sim 100$ larger than the average
density of the current, $\overline{j}=\frac{I}{2\pi a^{2}}$, since $%
I_{1}\approx I$. This suggests that in the narrow layer at the edge of the
tokamak, at the onset of the $H$ mode regime where there is a strong sheared
poloidal rotation (\emph{i.e.} high vorticity), there is an accumulation of
the current density. This appears to be compatible with the experimental
observations (see Fig.9 of \cite{burrellcurrent}).

This contribution to the current density in the layer is very large and
depends on the vorticity $\omega =\partial v_{\theta }/\partial r$. This is
not unexpected since $\partial \omega /\partial t=\left( B/\rho \right)
\nabla _{\parallel }j_{\parallel }$ is the dynamics in the phase of the rise
of the laminar sheared flow in the $H$-mode layer \cite{hortondrift90}. This
current is not transitory and persists as long as there is no substantial
change of the vorticity $\omega $. We note that it is not directly connected
with the \emph{bootstrap} current since the latter is generated by the
gradients in the pedestal. The fact that at the edge in $H$ - mode there is
a vorticity-current layer have several consequences: the Resonant Magnetic
Perturbations (RMP) have difficulties to penetrate the shell of current; the
edge localized modes (ELM) should have a \textquotedblleft
tearing-mode\textquotedblright\ component \cite{bulanovsasorov78}.

\section{Conclusions}

This paper examines a connection between the vorticity, the current density
profile and the density $\left( \omega ,q,n\right) $ as suggested by a
frozen-in MHD invariant.

We have first briefly contrasted two classes of invariants useful in the
study of the plasma (and fluids): the topological invariants, which are
non-local; and Lagrangian and frozen-in invariants, which exist as local
constrains to the plasma dynamics. The latter can be very useful in
applications, as are the conservation of energy or angular momentum. We have
noted two difficulties, which still wait for solutions: (1) the dissipation
renders the invariance approximative; (2) and, the conversion from
Lagrangian to Eulerian requires certain characteristics of the turbulence,
essentially the ergodicity that makes that an element of plasma carrying its
"charge" (Lagrangian invariant) can come close to any other element of
plasma, in a finite region. This problem is known in the evolution to
coherent asymptotic states of the $2D$ Euler fluid \cite{robertsommeria}.

We have mentioned the derivation of Lagrangian and frozen-in invariants, to
underline the particular distinction between the approach based on MHD
equations and respectively the turbulent equipartition. We have examined the
frozen-in MHD invariant derived by Sagdeev, Moiseev, Tur and Yanovskii. The
version for ions is particularly useful for the effects of changes of the
plasma sheared velocity. Especially the fast changes, like the onset of
sheard plasma rotation (vorticity) must be accompanied by a change in the
other plasma variables: safety factor $q$ and density $n$. We have mentioned
that the connection between $\left( \omega ,q,n\right) $ imposed by the SMTY
invariant suggests to consider all sources of sheared rotation (in
particular those able to form internal transport barrier) as connected with
the $q$-profile. ICRH, ECRH become possible sources of changes on the
profile of the current density. Conversely, LHCD, by modifying the current
profile, can induce transient sheared rotation of plasma in a finite region.
A very high vorticity concentrated around the magnetic axis can expell the
current density from the center producing a current hole.

One of the objectives was to look for consequences of the invariance when
vorticity is changed by the ionization-induced plasma rotation (at pellet
fuelling, or gas-puff or impurity seeding). \ We have illustrated this by
two examples: an effect contributing to the reversed-$q$ profile in the
plasma core and the effect of accumulation of density of current in the $H$%
-mode rotation layer at the edge.

Previous experimental observations that appear to be compatible with the connection expressed by this invariant must be re-examined: pellet enhanced performance, reversed shear (and enhanced reversed shear), current hole, etc. Numerical work will certainly be very helpful.

\bigskip

\textbf{Acknowledgments} This work has been partly supported by the Romanian
Ministry of National Education, through the Grant 1EU-10.


\providecommand{\noopsort}[1]{}\providecommand{\singleletter}[1]{#1}%

\end{document}